# Dynamic axion field in the magnetoelectric antiferromagnet chromia

Junlei Wang[1], Chao Lei[2], Allan H. MacDonald[2] and Christian Binek[1]

[1]Department of Physics and Astronomy, University of Nebraska-Lincoln, Lincoln, NE 68588

[2]Department of Physics, University of Texas at Austin, Austin, TX 78712

**Abstract**

Chromia is a magnetoelectric insulator whose bulk magnetoelectric susceptibility contains a non-zero pseudoscalar component, like that present in magnetized topological insulators. We reveal the dynamic axion field of chromia by measuring the temperature dependence of its non-linear magnetoelectric response using a lock-in technique with an AC electric field stimulus. The electric field dependence, magnetic field dependence, and temperature behavior of the second harmonic lock-in signal are all in agreement with theoretical expectations. Our results demonstrate that chromia's magnetoelectric response couples strongly to its low-energy quantum and thermal spin fluctuations.

**Introduction**

Maxwell's field equations are the Euler-Lagrange equations of the Maxwell action, whose Lagrangian density is given by $L_{EM} = \frac{1}{2}\left(\varepsilon \vec{E}^2 - \frac{1}{\mu}\vec{B}^2\right) - \rho\Phi + \vec{j}\cdot\vec{A}$, where $\vec{E}$ and $\vec{B}$ are electric and magnetic fields, $\Phi$ and $\vec{A}$ are scalar and vector potentials, and $\varepsilon$ and $\mu$ are frequency-dependent material parameters that account for linear coupling to matter. In magnetoelectric materials $L_{EM}$ contains an additional term $L_\theta \propto \theta \vec{E}\cdot\vec{B}$ which couples electric and magnetic fields. In a uniform magnetoelectric material the parameter $\theta$ [1,2] is time and position independent and Maxwell's equations are unchanged by $L_\theta$, except at surfaces and interfaces. The Maxwell action, including the magnetoelectric (ME) term when present, accounts fully for the linear response of matter to electric and magnetic fields and is partially analogous [3,4] to the action of axion field models considered in particle physics. A richer analogy [3,4] is realized when low-energy degrees of freedom present in the matter can be described by elevating $\theta$ from a materials parameter to a dynamic quantum field. In this paper we argue that chromia contains a dynamic axion field and identify its presence experimentally by observing a characteristic [3] temperature-dependent non-linear ME response.

Pseudoscalar $\theta \vec{E}\cdot\vec{B}$ magnetoelectric coupling is a characteristic low-temperature property of three-dimensional topological insulators (TIs) with weakly-broken [5] time-reversal-symmetry, which have an axion field value $\theta = \pi$ associated with their topologically nontrivial electronic band structure and topologically protected surface states [6,7]. The description of the ME response of magnetized TIs (the topological magnetoelectric effect, TME) in the framework of axion electrodynamics with a non-zero $\theta$ is equivalent [8] to conventional Maxwell's field equations that add a surface Hall conductivity or a bulk pseudoscalar ME susceptibility to the Maxwell equation constituent relations. The pseudoscalar nature of the topological ME susceptibility is rather unusual and gives rise to an isotropic ME effect in which the induced magnetization (polarization) and applied electric (magnetic) field are collinear. Although no known bulk cubic ME crystal has a pure pseudoscalar ME susceptibility [9], chromia and many other common ME materials do have large pseudoscalar ME response components that allow them to be used as practical platforms for axion electrodynamics.

There are several theoretical proposals for the realization of dynamic axion fields in condensed matter systems which could be detected via magneto-optical non-linearity, including proposals involving magnetized TI superlattices [3,4,10]. To the best of the authors' knowledge, however, no experimental confirmation of these proposals has been achieved. Detection of axion dynamics presents a variety of challenges, including signal-to-noise ratios that are expected to be low. This is especially true for experiments that utilize torque to detune the eigenfrequency of an oscillating cantilever with a magnetic TI tip [3]. In this approach small periodically varying torques originate from simultaneous exposure of the magnetic TI to an AC electric and a DC magnetic field. The induced sinusoidal magnetization is expected to experience torque in the presence of an applied magnetic field which would shift the eigenfrequency of the cantilever. In such an experiment, the dynamic axion field would be revealed through higher harmonic components in the frequency shift. As outlined in detail below, we perform an analogous optical experiment avoiding complications associated with a mechanical setup. We choose to study chromia because it has a high antiferromagnetic (AFM) ordering temperature, $T_N \sim 307$ K [11] compared to magnetic ordering temperatures around 15 K [12,13] in magnetic TIs such as Cr-doped $(Bi, Sb)_2 Te_3$, circumventing the need for low temperature setups. These two advantages allow us to measure the dynamic axion field using a rather straight-forward Faraday setup with an AC electric field that induces temporally periodic magnetization that we detect optically. The non-linear magneto-optical response can be singled out through a second harmonic signal with characteristic electric field, magnetic field, and temperature dependencies.

**Pseudoscalar axion response and dynamic axion field in chromia**

Although pseudoscalar (axion) ME response can be counterintuitive, as has been emphasized by Post [14], non-zero axion contributions are [9] generic in ME materials and do not require a topologically non-trivial electronic band structure [9,15]. In the ME antiferromagnet $Cr_2O_3$ (chromia) the presence of a non-zero pseudoscalar

$$\theta = \frac{1}{3}(2\alpha_\perp + \alpha_\parallel) \tag{1}$$

is an established experimental fact [2,9,16,17]. Ordinary ME materials therefore provide the ME response required by Wilczek [1] when he pointed out that condensed matter systems can in principle mimic axion fields [1]. Here $\alpha_\perp$ and $\alpha_\parallel$ are the perpendicular and parallel components of the ME susceptibility tensor $\underline{\underline{\alpha}}$ which relates magnetization and electric field according to $\mu_0 \vec{M} = \underline{\underline{\alpha}} \vec{E}$, and polarization and magnetic field correspondingly [18,19]. Chromia's ME susceptibility tensor is diagonal and

$$\underline{\underline{\alpha}} = \begin{pmatrix} \alpha_\perp & 0 & 0 \\ 0 & \alpha_\perp & 0 \\ 0 & 0 & \alpha_\parallel \end{pmatrix} = \frac{1}{3}(\alpha_\perp - \alpha_\parallel) \begin{pmatrix} 1 & 0 & 0 \\ 0 & 1 & 0 \\ 0 & 0 & -2 \end{pmatrix} + \theta \begin{pmatrix} 1 & 0 & 0 \\ 0 & 1 & 0 \\ 0 & 0 & 1 \end{pmatrix} \tag{2}$$

can be decomposed into a trace free component and a pseudoscalar or axion piece $\theta \delta_b^a$ [9]. The axion piece and its temperature dependence are well-established for chromia. In analogy to magnetic TIs, the magnetization of chromia couples to the AFM order parameter which acts [4] as a dynamic axion field. However, no experiments have been reported which use a trivial ME insulator as a testbed for the demonstration of a dynamic axion field.

**AC electric field induced Faraday rotation for detection of the dynamic axion field**

Central to our magneto-optical detection of the dynamic axion field is a recently established table-top setup which allows us to measure the AFM order parameter orientation in ME antiferromagnets in general, and in chromia in particular [20]. Fig. 1 shows the magneto-optical setup (for details see method). Here the interaction between the probing light and the sample gives rise to an electric Faraday effect at $T < T_N$, and therefore periodic rotation of the plane of polarization in response to a kHz electric AC field applied along chromia's *c*-axis. We view the Faraday signal as a proxy for magnetization. The *E*-field induced Faraday rotation has two components that superimpose linearly, one proportional to the ME response and one to a pseudo Stark effect [20,21] that is proportional to the AFM order parameter. Coefficients $p(\lambda)$ and $q(\lambda)$ which weigh the contributions in linear superposition, depend strongly on the wavelength, $\lambda$, of the probing light [20]. This property makes it possible to select pure response from either of the two components. We select a probing wavelength of $\lambda = 940$ nm for which the Faraday response originates primarily from the ME susceptibility, while at the same time absorption is weak enough to maintain a beneficial signal-to-noise ratio. A magnetic DC field of variable strength can be applied simultaneously along the *c*-axis with the AC electric field. The resulting photovoltaic signal is detected by phase-sensitive lock-in technique. Its second harmonic contains the information about the dynamic axion response.

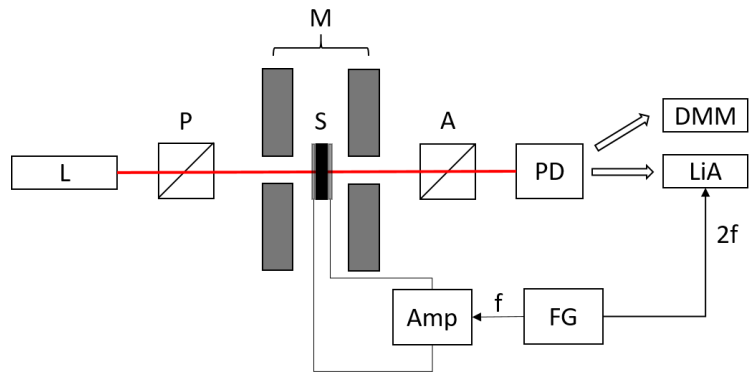

**Fig. 1**: Magneto-optical setup for measurement of the dynamic axion field via electric field induced Faraday rotation. L: laser; P: polarizer; M: electro-magnet; S: sample inside UHV chamber with electric contact; A: analyzer; PD: photodetector; DMM: digital Multimeter; LiA: lock-in amplifier; Amp: AC amplifier; FG: function generator.

Conceptually, our magneto-optical setup is reminiscent of the mechanical cantilever experiment proposed for TMIs [3]. In the case of a static axion field, for example one that corresponds microscopically to electric-field-induced canting of the host antiferromagnet's moments, one expects a time dependent contribution to the *E*-field induced Faraday rotation which rigidly follows the sinusoidal AC excitation. However, when the axion response is due electric-field-induced coupling between total magnetization and dynamic low-energy degrees of freedom in the antiferromagnet, the Faraday response is expected to include a prominent temperature-dependent non-linear component signaled by higher order harmonics. In order to discriminate the dynamic axion field from ordinary ME non-linearities a detailed analysis of the electric field dependence, magnetic field dependence, and temperature behavior of the second harmonic lock-in signal is required.

Chromia is an antiferromagnetic insulator with low energy S=3/2 chromium spin and lattice degrees-of-freedom. Its ME response is due mainly [22-24] to changes in interactions in the spin system that occur when the chromium ions move in response to an applied electric field. Below we assume that dynamic fluctuations of the lattice do not play a role and describe the system in terms of quantum spins alone. Chromia's spin-Hamiltonian can be expressed in the form [23,24]:

$$H = -\frac{1}{4}\sum_{X'X\mathbf{k}} J_{X'X}(\mathbf{k})\, \vec{S}_{X'}(-\mathbf{k}) \cdot \vec{S}_X(\mathbf{k}) \qquad (3)$$

where $\vec{S}$ is a spin-operator, $\mathbf{k}$ is a wave-vector in chromia's rhombohedral Brillouin-zone and $X', X = G, F, A, C$ label symmetry adapted combinations of the four spins per unit cell. Below we focus on coupling between the spin that orders at $\vec{E} = 0$, $\vec{S}_G = \vec{S}_1 - \vec{S}_2 + \vec{S}_3 - \vec{S}_4$ and the total spin, which influences macroscopic electrodynamics $\vec{S}_F = \vec{S}_1 + \vec{S}_2 + \vec{S}_3 + \vec{S}_4$. When chormia's exchange interactions are expanded in powers of $\vec{E}$, the leading term in $J_{FG}(\boldsymbol{k} = 0)$ is linear. Its presence identifies $\vec{S}_G$ as a dynamic axion field; macroscopic magnetization ($\langle\vec{S}_F\rangle$) is induced by an electric field because $\vec{S}_F$ is coupled to $\vec{S}_G$. Both $\vec{S}_G$ and the magnetization $\vec{S}_F$ are dynamic fields and have correlated quantum dynamics.

Chromia's non-linear magnetoelectric response can be described qualitatively using a mean-field approximation for the quantum spin-Hamiltonian. We define $\langle\vec{S}_X(\boldsymbol{k} = 0)\rangle = S_X$. Assuming for simplicity that $\vec{S}_A$ and $\vec{S}_C$ can be neglected, it follows [See Supplementary Material] that the antiferromagnetic order parameter can be expressed as $g = 2\left[B_{\frac{3}{2}}(\beta(h_g + h_f)) + B_{\frac{3}{2}}(\beta(h_g - h_f))\right]$, and the ferromagnetic order parameter as $f = 2\left[B_{\frac{3}{2}}(\beta(h_g + h_f)) - B_{\frac{3}{2}}(\beta(h_g - h_f))\right]$, where $h_g = \frac{1}{4}[J_{GG} S^2 g + J_{GF}S^2 f]$, $h_f = \frac{1}{4}[J_{FF} S^2 f + J_{FG}S^2 g + HS]$ and the exchange coupling constants are understood to be evaluated at wave-vector $\boldsymbol{k} = 0$. We show below that these equations are able to explain our observations.

**Experimental results**

Fig. 2 shows an example of the measured (circles) second harmonic lock-in signal $V_{2\omega}$ vs. $E_0$. The data shown in Fig. 2 were taken at temperatures $T = 92$ K, 128 K, 148 K and 192 K, with static magnetic field $B_0$=110 mT, and electric field $E(t) = E_0 \sin \omega t$ with AC frequency $\omega/2\pi = 3135$ Hz. The lines are best fits of the functional form $V_{2\omega} = P_1 E_0^2$ to the corresponding data set.

The quadratic $E_0$ – dependence verifies that at this frequency $V_{2\omega}$ is due to adiabatic non-linear response of magnetization to electric field, *i.e.* it is a measurement of $\frac{d^2 f}{d E_0^2}$.

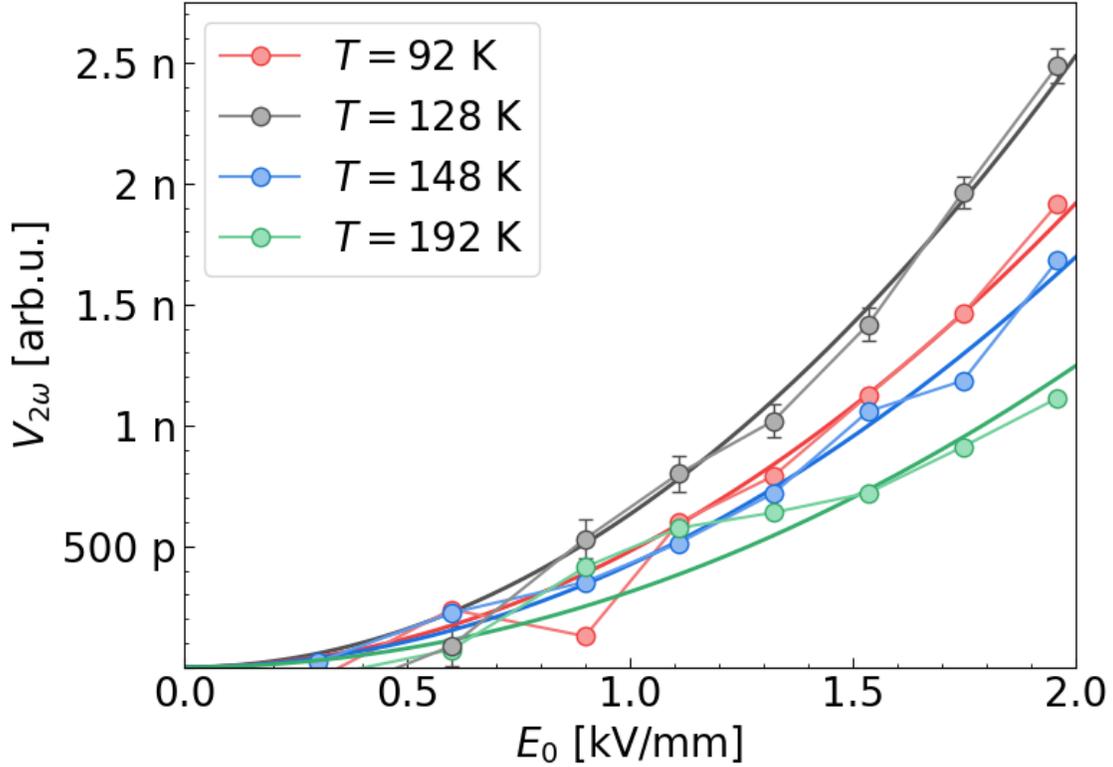

**Fig. 2**: Second harmonic signal $V_{2\omega}$ (circles of various colors) versus amplitude $E_0$ of applied AC electric field of $f$=3135 Hz. Data are taken at $T$=92 K (red), 128 K (grey), 148 K (blue) and 192 K (green), respectively, in the presence of a static $B_0$-field of 110 mT. Lines of same colors are a single parameter best fit of $V_{2\omega} = P_1 E_0^2$ of the corresponding data. The error bars on the grey data show one standard error.

Fig. 3 shows three data sets for $V_{2\omega}$ vs. $B_0$ measured at $T$=129 K and with $\omega/2\pi = 3135$ Hz. Each data point is obtained by averaging a sequence of 500 measurements at the same $B_0$ and $E_0$ magnitudes. The second harmonic signal in a ME material is an odd function of both electric and magnetic external fields, i.e., the sum of the power of $E_0$ and $B_0$ is always odd. The presence of a finite second harmonic signal in Fig.3 at $B_0 = 0$ signals the presence of either weak residual magnetism, e.g., due to imperfect moment compensation at surfaces or bulk defects, or an associated small residual internal electric field $E_0$: $\frac{\partial^2 f}{\partial E_0^2} = \frac{\partial^3 f}{\partial E_0^3}\bigg| E_0 + \frac{\partial^3 f}{\partial E_0^2 B_0}\bigg| B_0$. The linear dependence of $V_{2\omega}$ on $B_0$ is apparent in Fig.3 and has the same value in a series of runs. The ill-defined shifts with respect to each other are experimental artifacts mainly due to the smallness of the signal and are within the noise of the measurement. The observed linear dependence on $B_0$ in concert with the pronounced $E_0^2$-dependence of $V_{2\omega}$ make a strong case for a non-linear ME response $V_{2\omega} \propto E_0^2 B_0$.

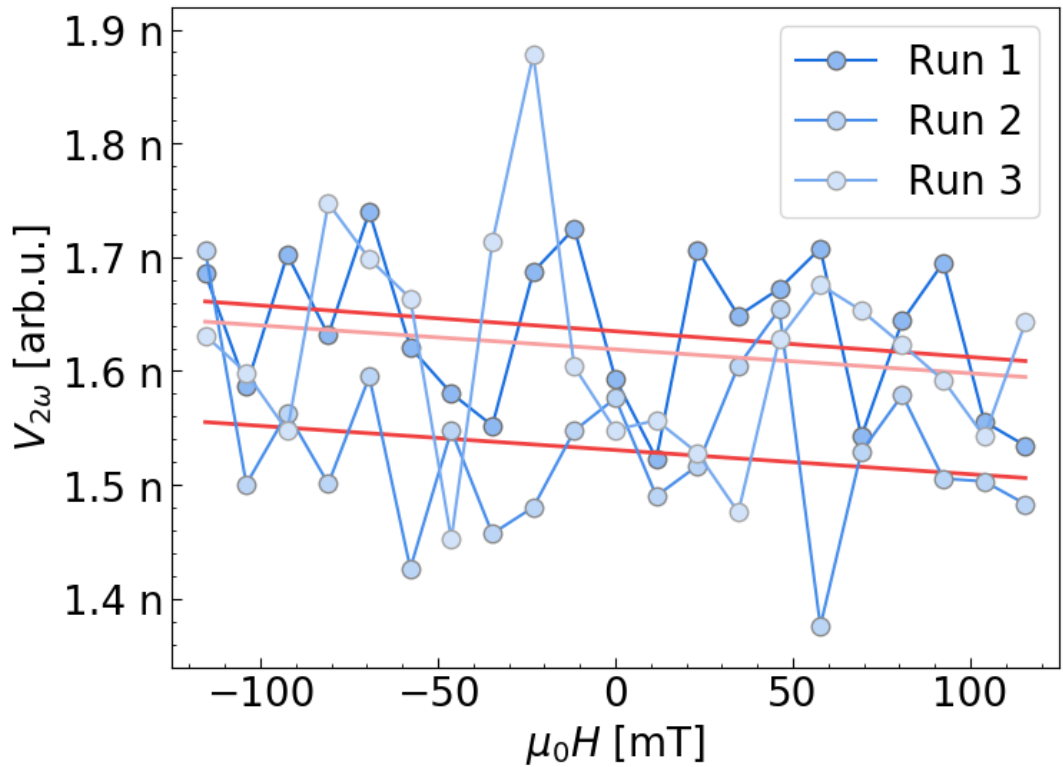

**Fig. 3**: Second harmonic signal $V_{2\omega}$ vs. $B_0$ measured in subsequent runs at $T$=129 K. The applied AC electric field has a frequency of $f = 3135$ Hz and amplitude of $E_0 = 1.6$ kV/mm. Lines are linear best fits.

**Temperature dependence of the dynamic axion signature**

The linear ME response $\alpha$ of chromia is [24] proportional to the antiferromagnetic order parameter $g$, and to the response $\chi$ of the average spin polarization $f$ to exchange or external fields. (See the supplementary material for further details.) Since $\chi$ is suppressed at low temperatures when the spins are locked in the antiferromagnetic ground state arrangement and $g$ vanishes above the critical temperature, $\alpha(T)$ [25] has a peak value at $T \sim 0.85\, T_N \sim 260$K. Higher order ME response can in principle arise from canting or non-linear dependence of exchange interactions on electric field, or from quantum and thermal fluctuations of the antiferromagnetic order parameter acting as a dynamic axion field.

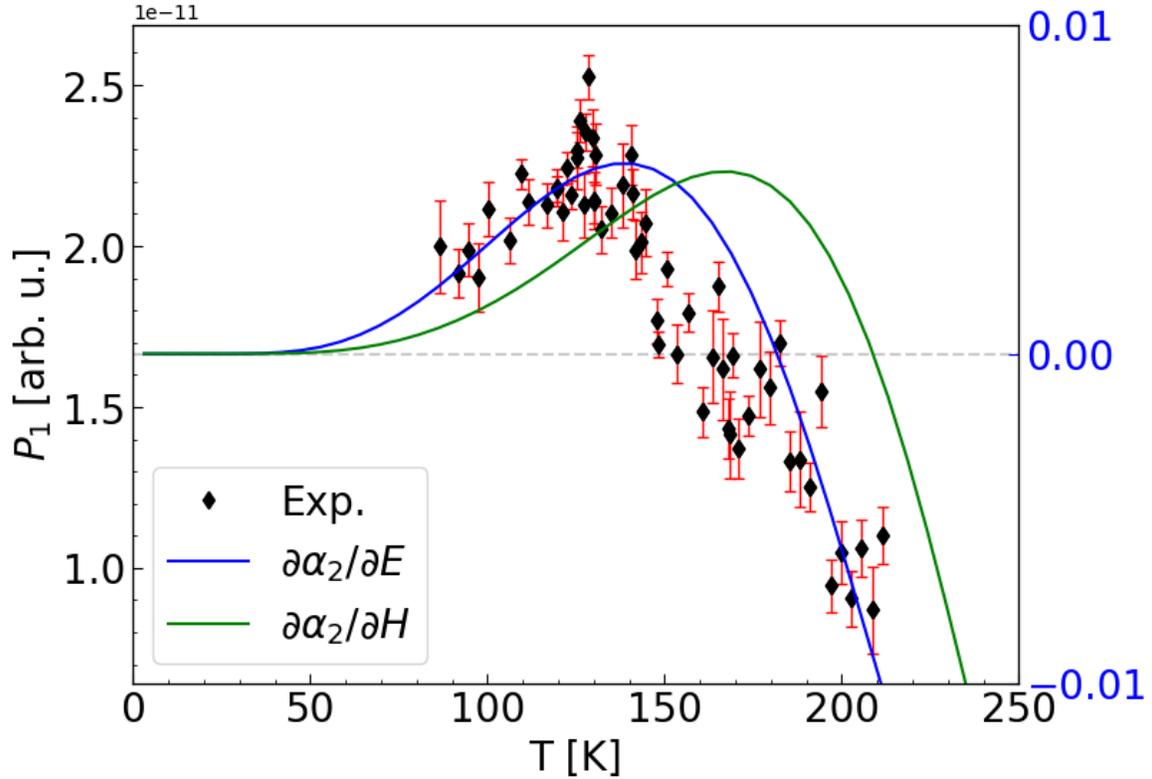

**Fig. 4**: $T$-dependence of the non-linear magnetoelectric response $\alpha_2$ of chromia. Each point $(P_1, T)$ is determined from a fit of the electric-field dependence of the non-linear signal at fixed $T$ to $V_{2\omega} = P_1 E_0^2$. The solid lines are mean-field theory calculations of the temperature dependence of $\alpha_2$ at weak electric field (blue) and at weak magnetic field (green). The non-linear response is peaked at a lower temperature than the linear response. The three curves are scaled to have a common peak value. The theoretical non-linear response functions vanish for temperature $T \to 0$ because the mean-field theory neglects quantum fluctuations in the antiferromagnetic ground state.

In Fig. 4 we compare our observations with theoretical mean-field ME response coefficients calculated assuming strictly linear dependence of $J_{FG}$ on electric field. Each experimental point $(P_1, T)$ is determined from a fit of the isotherm $V_{2\omega} = P_1 E_0^2$ for various electric field amplitudes. The theory curves show mean-field estimates for the temperature dependence of second order non-linear ME response $\alpha_2$ at weak electric field (blue) and at weak magnetic field (green). The lower temperature of the peak in the non-linear $P_1$ signal, at $T \sim 0.4\, T_N \sim 130K$, compared to the peak in the $\alpha$ signal, is accurately reproduced, directly establishing coupling between magnetization and a fluctuating quantum order parameter as the source of ME response of chromia and providing an accessible physically realizable dynamic axion field.

**Conclusions**

Building on the fact that a small axion field exists in the trivial magnetoelectric insulator chromia we demonstrate its dynamic character by measuring the electric field induced Faraday effect, i.e., the rotation of the polarization plane of transmitted linearly polarized light in response to an AC electric field. The latter is applied along chromia's *c*-axis simultaneously with a stationary magnetic field. The presence of a second harmonic signal in the detected light intensity is a hallmark of the dynamic axion field. The quadratic electric field dependence, linear magnetic field dependence, and non-trivial temperature dependence of this second harmonic signal are in agreement with theoretical expectations. Our results thus demonstrate the presence of a dynamic axion field in a trivial insulator with broken space and time inversion.

**Acknowledgement**

We gratefully acknowledge financial support by the Army Research Office through the MURI program under Grant Number W911NF-16-1-0472. This research was supported in part by the National Science Foundation, through the Nebraska Materials Research Science and Engineering Center (MRSEC) Grant No. DMR-1420645 and by the Welch Foundation through grant Welch TBF1473. The research is performed in part in the Nebraska Nanoscale Facility, which is supported by the National Science Foundation under Grant No. NNCI: 1542182